\documentclass{article}
\usepackage[a4paper, margin=1.5in]{geometry} 

\usepackage{graphicx} 
\usepackage{amsmath} 
\usepackage{caption}

\usepackage[backend=bibtex, style=numeric]{biblatex}

\title{\textbf{Dynamic Fraud Proof}}
\author{Gabriele Picco, Andrea Fortugno}
\date{www.magicblock.gg}

\begin{document}

\maketitle

\begin{abstract} 

In this paper, we present a novel fraud-proof mechanism that achieves fast finality and, when combined with optimistic execution, enables real-time transaction processing. State-of-the-art optimistic rollups typically adopt a 7-day challenge window, during which any honest party can raise a challenge in case of fraud. We propose a new assert/challenge construction called "Dynamic Fraud Proofs" that achieves sub-second finality in ideal scenarios, while dynamically delaying settlement in the event of fraud detection and challenge resolution. The system relies on 1) a dynamic challenge period and 2) a configurable number of randomly selected verifier nodes who must interactively approve a state commitment without raising a challenge. If these conditions are not met, the state is not finalized, and the challenge period and approval criteria are dynamically adjusted.
We provide a detailed analysis of the system's design, explaining how it maintains the assumption of a single honest node and addresses censorship attacks by inverting the traditional challenge process. Additionally, we formalize the system's probabilistic security model and discuss how bonding, incentives, and slashing mechanisms can encourage honest behavior, thereby increasing the likelihood of fast settlement in ideal scenarios.

\end{abstract}

\section{Introduction}

Ephemeral Rollups, introduced in \cite{picco2023ephemeral}, are a scaling solution designed to enable elastic throughput, low latency, and enhanced capabilities for the base Solana Virtual Machine (SVM) layer. The key innovation of Ephemeral Rollups is that state, logic (blockchain programs), and users remain on the base layer, while the tooling and infrastructure are fully compatible with the virtual machine specifications. Execution can be delegated to one or more parallel auxiliary layers that adhere to the same virtual machine standards. This approach prevents ecosystem fragmentation by avoiding sharding or permanent migration of state, users, and programs to a new blockchain or layer.

This paper presents the mechanism used to validate off-chain computation performed in Ephemeral Rollups (ERs) using a fraud-proof design that aims to (1) dynamically minimize the challenge period under ideal conditions, and (2) address the issues that arise from a shortened challenge period. 

A similar idea has been explored in the context of the Ethereum Virtual Machine (EVM) in \cite{wweek}, where the challenge period changes dynamically to improve finality. Our approach extends this concept to Solana by leveraging the inherent parallelism and peculiarities of the SVM. 

ER execution and state changes are managed optimistically, with updates propagated to clients, as soon as they are processed, while settlement is subject to a challenge process for verifying execution and state differences. This is commonly referred to as the "challenge period" in optimistic setups. The fraud-proof mechanism, termed dynamic fraud proof, employs a dynamic and auto-adaptable challenge period. This period can vary based on several factors, including the value at stake in the rollup session, the participants in the validation process, and the economic bonds involved. Under honest conditions, the challenge period—and thus finality—can be nearly real-time, extending only when optimal conditions are not met.

The following section provides background information essential for understanding the subsequent discussion on the design, architectural choices, security implications, and optimizations that ensure the system's practicality while maintaining a level of security effectively equivalent to that of the base layer.

\cite{wweek}

\subsection{SVM runtime}
\label{ss:svm-runtime}

The system implementation and the discussion of the following sections relies on some unique properties of the Solana Virtual Machine (SVM), which we highlight below:

\begin{itemize} 

\item Unlike most blockchains that organize state data into tree structures (e.g., Merkle trees, Verkle trees), the SVM does not maintain a committed global state tree. Solana's approach allows transactions that do not modify the same state (stored in accounts) to execute in parallel. 

\item The SVM uses an account model. Accounts can store lamports and bytes. They can be either executable (binaries of the smart contracts) or non-executable. Logic and state are natively separated, and each smart contract stores and modifies the state held in multiple owned accounts (PDAs).

\item Transaction on Solana declare upfront the state they access, specifying accounts as readable or writable. This allows using the mechanism described in \cite{picco2023ephemeral}, and transparent modification of the state in parallel on the base SVM and the Ephemeral Rollups sessions. 

\item The fast block time allows for commitment and finalization of the state on the base layer, which could be temporarily advanced in an Ephemeral Rollup instance.
The finalization of the state is subject to the challenge period, but can be relatively short as explained in [2].

\end{itemize}

\subsection{Proof Systems}
\label{ss:proof-system}

As a brief recap, the two primary approaches for verifying the correctness of off-chain state transitions are fraud proofs, used by optimistic rollups \cite{DBLP:journals/corr/abs-1904-06441}, and validity proofs, used by validity rollups \cite{proofs}.

\begin{itemize}
    \item Fraud proofs: These proofs assume transactions batches are correct unless proven fraudulent within a challenge period. 
    \item Validity proofs: This approach requires each state transition to be accompanied by a cryptographic proof of its correctness. 
\end{itemize}

It’s important to note that proof systems (e.g., STARK \cite{BenSasson2018ScalableTA}, SNARK \cite{chen2022review}) can be used for both fraud proofs and validity proofs. Validity rollups adopt full proof validation, i.e., they produce a cryptographic proof for every state change, validating it before settling the state. In contrast, interactive fraud proving (optimistic rollups) does so only when a challenge is raised. Mixed schemas, such as “naysayer proofs” \cite{cryptoeprint:2023/1472} exist where a verifier optimistically accepts a submitted proof without verifying its correctness, allowing an observer to check the proof off-chain and eventually raise a challenge.

In the following sections, we discuss how to achieve fast finality in an optimistic setup using a dynamic challenge window. While the Dynamic Fraud Proof (DFP) approach is broadly applicable, we contextualize it within the ephemeral rollups design, summarized in the next section. The focus of this discussion is on dynamically optimizing the challenge period and how the system is designed to efficiently detect fraud. Dispute resolution, though beyond the scope of this discussion, can leverage techniques such as zkVM or mechanisms similar to Arbitrum’s BoLD resolution, as described in \cite{alvarez2024boldfastcheapdispute}.

\section{System Architecture}

\begin{figure*}[t!]
    \includegraphics[width=\textwidth]{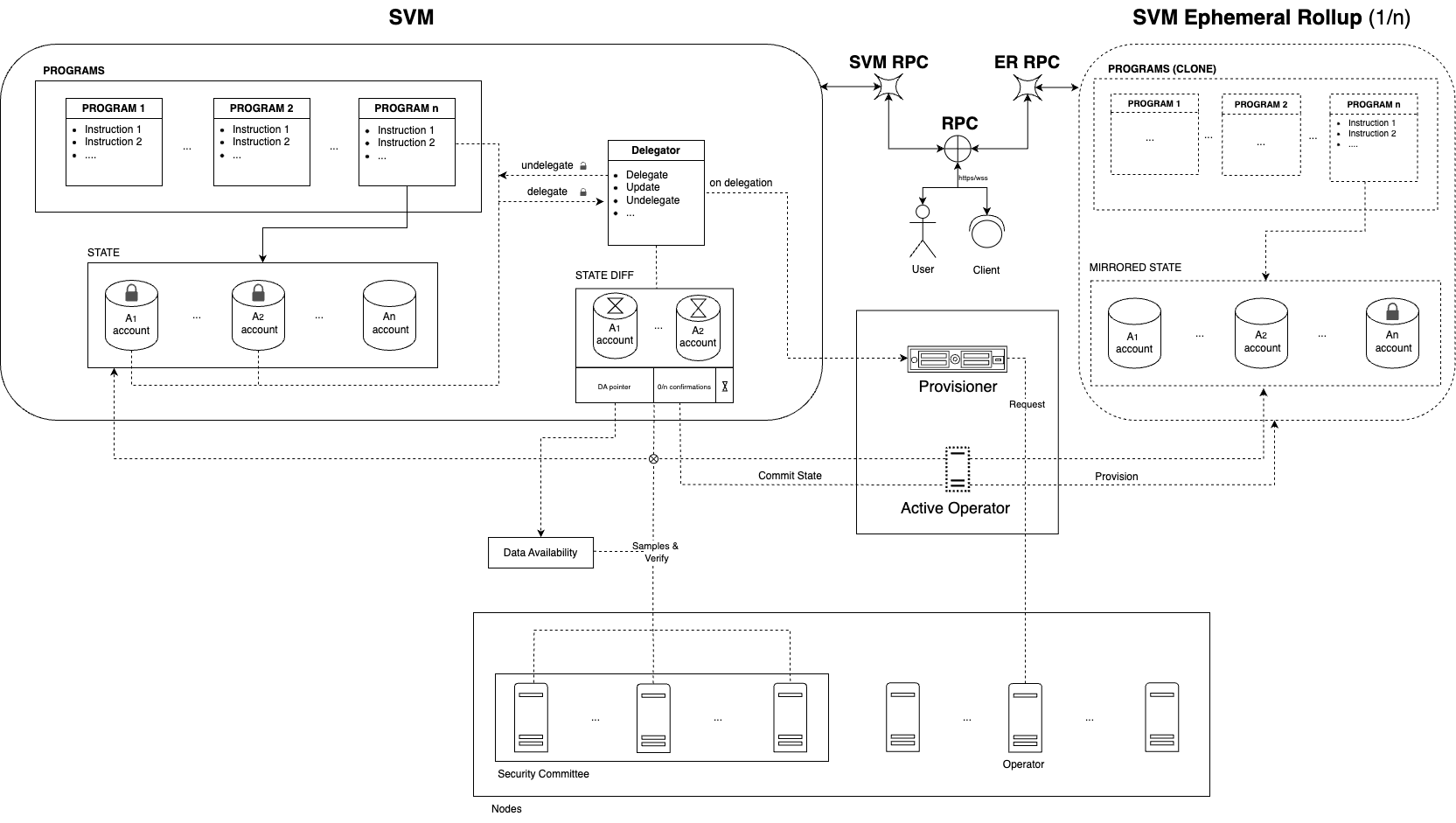}
    \caption{The system is comprised of several key components: The Delegation Program manages account locking/unlocking, state commitments, and finalization on the base SVM. The Operator is a node running an Ephemeral SVM session, while the Provisioner listens for delegation events, selects operators, and manages runtime provisioning. The Security Committee verifies execution and data availability, supported by the Data Availability Layer that stores transactions from the ephemeral session. An RPC Router offers a convenient interface for clients to access and modify state across both the base layer and Ephemeral Rollup sessions.}
    \label{fig1:arch}
\end{figure*}

For a more extensive description of the architecture, see \cite{picco2023ephemeral}. Below is a summary of the main components of the system described in the companion paper and illustrated in Figure \ref{fig1:arch}:

\begin{itemize}
    \item \textbf{Delegation Program}: Deployed on the base SVM, this program locks/unlocks accounts (trough native delegation), commits state changes, and is responsible for finalizing the state after the security conditions are satisfied.
    \item \textbf{Operator}: A node actively running an Ephemeral SVM session.
    \item \textbf{Provisioner}: Actively listens for delegation events, selects an operator, and guides the runtime provisioning process. The provisioner can be implemented as a smart contract, with selection based on a market of supply and demand where operators serve as session runtimes for delegated accounts.
    \item \textbf{Security Committee}: Nodes that verify execution and data availability.
    \item \textbf{Data Availability Layer}: Stores the transactions executed in the ephemeral session.
    \item \textbf{RPC Router}: Provides a convenient entry point for clients to transparently access and modify the state on both the base layer and the Ephemeral Rollups sessions.
\end{itemize}

\subsection{Execution}
\label{s:execution-and-fp}

Ephemeral Rollups enable off-chain computation by delegating state modifications and access to a layer with lower latency and reduced costs compared to the base layer. Moreover, the runtime environment can incorporate customizations, such as transaction scheduling. The following outlines a typical process that a program can adopt to facilitate the execution of transactions within an Ephemeral Rollup (ER) session: 

\begin{enumerate}
    \item A program interacts with the Delegation Program (DLP) to delegate state (i.e., accounts). This delegation includes metadata such as the maximum lifetime of the delegation, the update frequency on the base layer, and the target block time. The delegated accounts are locked on the base layer and can only be updated or undelegated when specific conditions are met.
    
    \item By monitoring the Delegation Program, the provisioner detects delegation events, selects an operator, and manages the runtime just-in-time provisioning based on the configuration. In an alternative implementation, operators can self-provision, thereby eliminating the need for this module.

    \item Clients execute transactions or retrieve data via RPC send their transactions to an RPC router. Based on the accounts involved in the transactions, the router forwards the transactions to the appropriate execution layer, either on the base layer or within one of the Ephemeral Rollup (ER) sessions. The router component serves as a convenience layer, but endpoint selection can also be performed directly by the client.

    \item Periodically, upon ER termination, or on demand (per account), the active operator commits the state to the base layer. This commitment on the base layer includes the new state, a pointer to the Data Availability (DA) record containing the transaction data executed in the ephemeral runtime, and a pointer to the delegation record, which specifies the conditions for approving and finalizing the state.
    
    \item The state is finalized using a dynamic fraud-proof mechanism as described in Section \ref{sec:fraud_proof}.
\end{enumerate}

\subsubsection{Atomic State Commitment \& Message Passing}
\label{ss:atomicity}

As discussed in Section \ref{s:execution-and-fp}, state commitments are made on a per-account basis with a configurable frequency, upon session termination or on demand. The latter can be triggered through Cross-Program Invocation (CPI) within transactions running in the Ephemeral Rollup (ER) session whenever the smart contract requires it. For example, when state needs to be surfaced to the base layer before executing a transaction on the base layer, e.g., a token swap or when claiming a reward.

The system also supports bundling multiple accounts for atomic commitment and finalization, ensuring that they are either atomically finalized or reverted. It is important to note that the protocol is general and treats state commitments as account diffs (i.e., byte arrays). This allows programs to use the same primitives to implement various message-passing or wrapping mechanisms between layers.

\section{Dynamic Fraud Proof}
\label{sec:fraud_proof}

We introduce a modified version of the canonical fraud proof system, termed Dynamic Fraud Proofs. The core mechanism follows the assert/challenge construction, wherein the asserted data must be subjected to a dispute period, allowing participants on the blockchain to enforce the state. A particular aspect of this design is its customizability of the challenge period (typically 7 days in other systems), with the goal of minimizing it to achieve faster settlement. This system retains the same single honest node assumption as traditional fraud proofs, wherein any honest party can issue a challenge. However, it relies on active challengers to address the free-rider problem; with a shorter challenge period (potentially on the order of seconds), the likelihood of fraud detection decreases. The design also incorporates an incentive structure by introducing payouts for active challengers, thereby increasing the probability of fraud detection and ensuring rapid settlement.

The process through which the state is updated from an Ephemeral Rollup (ER) is as follows:

\begin{enumerate}
    \item The active ER node submits a commitment of the state difference (diff) it intends to finalize.
    
    \item A challenge window is initiated, during which any participant can raise a challenge to identify an invalid state commitment. Challengers utilize the data availability from the ER session to verify the SVM execution.

    \item If a challenge is raised, a dispute resolution process, referred to as a dispute game, begins. The state cannot be finalized until the dispute is resolved.
    
    \item At the end of the challenge period, one of two outcomes may occur:

    \begin{enumerate}
        \item If a sufficient number of randomly selected active challengers (meeting a preconfigured threshold) have signed off on the state diff (i.e., have not raised a challenge), the state is settled.
        \item If an insufficient number of challengers have signed off on the state diff, the challenge window is extended.
    \end{enumerate}
\end{enumerate}

The optimistic approach and fraud proof mechanism share similarities with the assert/challenge constructions described in works such as \cite{Poon2017PlasmaS}, \cite{DBLP:journals/corr/abs-1904-06441} and \cite{Kalodner2018ArbitrumSP}. The primary difference lies in the granularity of what can be moved off-chain using the delegation mechanism described in Section \ref{s:execution-and-fp}. Each ER can process and commit state asynchronously to the base layer, with varying frequencies and internal block times. Moreover, state updates for different accounts can be proposed and finalized at different intervals or on demand, even within the same ER session. For a more in-depth discussion on commitment and how to achieve atomic commitment of multiple accounts, refer to Section \ref{ss:atomicity}.

\begin{figure*}[t!]
    \includegraphics[width=\textwidth]{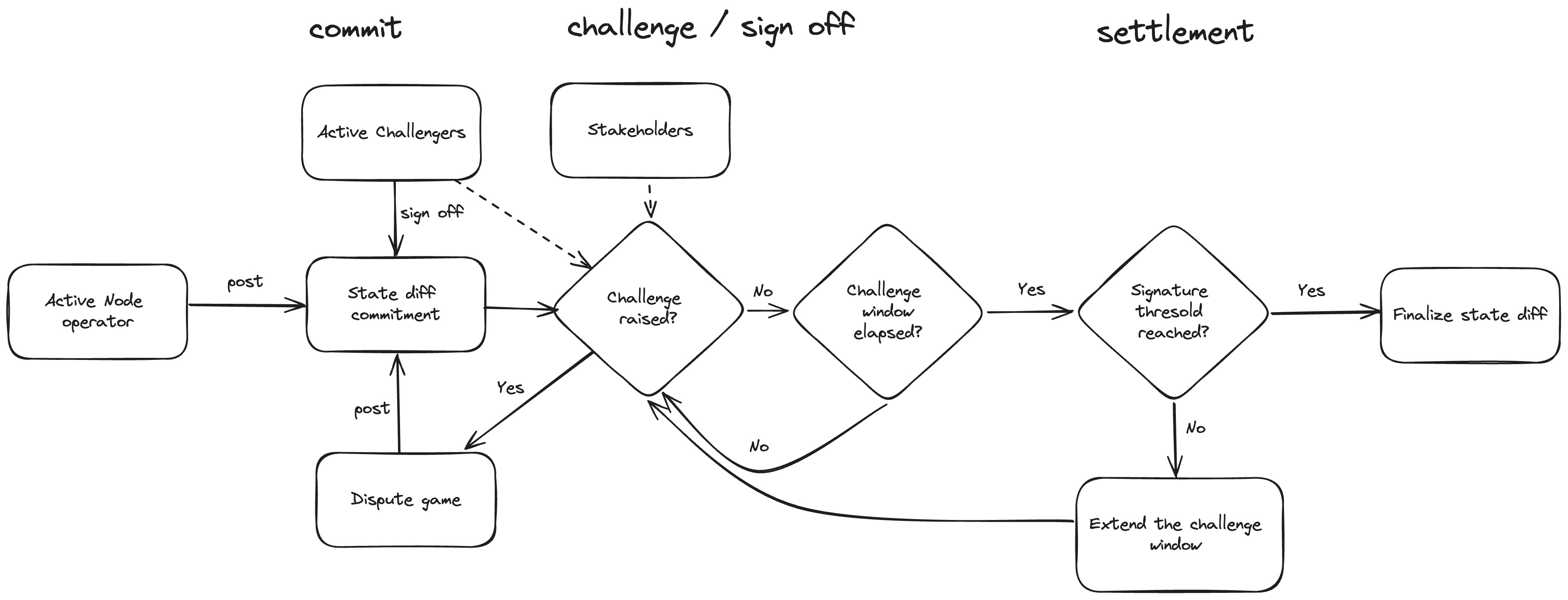}
    \caption{The process of state finalization in an ER involves multiple steps to ensure integrity and security. The ER node first submits a commitment of the state difference it intends to finalize. A challenge window is then initiated, allowing participants to raise challenges against any invalid state commitment. If a challenge is raised, a dispute resolution process begins, preventing the state from being finalized until the dispute is resolved. Depending on the outcome of the challenge period—whether enough challengers sign off on the state diff—the state is either settled or the challenge window is extended.}
    \label{fig2:multi-dimensional}
\end{figure*}

This design difference allows for sharding the state (i.e., accounts) of the same smart contract between the base layer and one or more Ephemeral Rollups. Transactions can be executed in parallel across the layers (base layer or ER sessions), depending on whether the accounts are delegated or not.

Fast finality is therefore essential to enable the swift transfer of state from the ER session to the base layer or across ER sessions (using the base layer for communication).

\section{Achieving Fast Finality}

While clients can operate on pre-confirmation when executing and reading state from ER sessions, which is sufficient for use cases such as gaming or an off-chain order book. Operations that require up-to-date state on the base layer must wait for the state to be finalized before executing transactions that depend on it. For example, a reward on the base layer that can be claimed after a certain state is reached, or an order book in the ER that then settles liquidity on the base layer, both require the finalization of state before executing the base layer transaction involving the delegated accounts.

It is therefore critical to achieve fast finality to maintain a high-quality user experience while carefully avoiding compromises to security. For a detailed discussion on trust assumptions, see Section \ref{ss:trust-assumptions}.

The finality of the state is constrained by the challenge period, which, in the case of ER, is customizable as described in Section \ref{s:execution-and-fp}, meaning the finality itself is also customizable.

Several studies have analyzed and discussed approaches to reducing the challenge period, as well as the implementation of dynamic challenge periods, such as those explored in \cite{why-7-days} and \cite{threesigma-challenge-period}. In \cite{offchain-optimizing-challenge-period}, the authors calculate the optimal theoretical challenge period under various assumptions, resulting in 1081 blocks, equivalent to 4.5 hours with a 15-second block time. On Solana, where the current block time is 400 milliseconds, this would be approximately 7.21 minutes. Although this duration falls within the range of minutes, it remains excessively long for interactive use cases, where users are actively waiting before proceeding with the next transaction.

\subsection{Short Challenge Period}

It is well known that a longer challenge period enhances security, as it makes censorship attacks on the chain more difficult. A malicious actor might attempt an attack by making an invalid assertion about a state transition and then censoring any attempts to challenge that assertion until the challenge period ends. The longer the challenge period, the less likely such an attack is to succeed.

The primary concept of the dynamic fraud proof mechanism is to invert the traditional challenge process. A configurable number of random nodes are required to sign a state commitment, thereby approving the state and not raising a challenge. As illustrated in Figure \ref{fig2:multi-dimensional}, if the threshold for approval is not met when the challenge period expires, the state cannot be finalized, and the challenge period is extended. 

This mechanism ensures two key outcomes: (1) there is a guarantee that a required number of parties have approved the proposed state, and (2) if someone is attempting to halt the network, the state will not be finalized. With these assumptions, the design allows for extremely short challenge periods, potentially in the range of seconds or sub-seconds (assuming active challengers are executing and verifying state transitions in parallel) under optimal conditions, with a mechanism to automatically extend the challenge period.

The challenge period can be extended according to an exponential function, with the threshold for active challengers being reduced accordingly.

\subsubsection{Extension of the Challenge Window}
To formalize the challenge period extension mechanism, let's define the challenge window time at step \( n \) as \( t_n \), starting with an initial time \( t_0 \) and growing exponentially with a factor \( r_t \):

\[
t_n = t_0 \cdot r_t^n
\]

where:
\begin{itemize}
    \item \( t_0 \) is the initial challenge window time (e.g., \( t_0 = 500 \) ms),
    \item \( r_t \) is the exponential growth factor for the time (e.g., \( r_t = 4 \) to represent a 4x increase from 500 ms to 2 seconds).
\end{itemize}

For example, with \( t_0 = 500 \) ms and \( r_t = 4 \), the sequence would be:

\[
t_0 = 500 \, \text{ms}, \quad t_1 = 2 \, \text{seconds}, \quad \text{and} \quad t_{10} \approx 6 \, \text{days}.
\]

\subsubsection{Decrease in the Number of Active Challengers}

Similarly, we can define the number of required active challengers at step \( n \) as \( c_n \), starting with an initial number \( c_0 \) and decreasing exponentially with a decay factor \( r_c \):

\[
c_n = c_0 \cdot r_c^n
\]

where:
\begin{itemize}
    \item \( c_0 \) is the initial number of challengers (e.g., \( c_0 = 100 \)),
    \item \( r_c \) is the exponential decay factor for the number of challengers (e.g., \( r_c = 0.7 \) to represent a 30\% decrease).
\end{itemize}

For example, with \( c_0 = 100 \) initial required active challengers and \( r_c = 0.7 \), the sequence would be:

\[
c_0 = 100, \quad c_1 = 70, \quad \text{and} \quad c_{10} \approx 2
\]

Note that the system allows anyone to submit a challenge. Active node challengers are required to counterbalance the short challenge period; however, when the challenge period is extended to 7 or more days, the system operates similarly to traditional fraud-proof mechanisms. In such cases, the number of active challengers can be zero, relying on at least one participant (anyone) to submit a challenge if necessary.

\subsection{Trust Model Formalization}
\label{ss:trust-assumptions}

In this section, we present a simplified formalization of the probability of raising a challenge in an ER, in order to discuss the likelihood of not achieving the ideal finality scenario and to highlight the effects of economic bonding, slashing, the number of nodes, and hyperparameters that can be tuned in different ER systems to achieve the desired security level.

\vspace{1em}

Let's define:

\vspace{1em}

\textbf{Event of Raising a Challenge}: Let \( E \) be the event of a challenge being raised against a committed state.

\vspace{1em}

\textbf{Probability of a Challenge}: Let \( P(E) \) represent the probability of raising a challenge.

\vspace{1em}

Where the factors influencing \( P(E) \) are as follows:

\begin{itemize}
    \item \textbf{Challenge Window \( (T) \)}: The duration during which a challenge can be raised. Let \( P(T) \) be the probability that a node detects fraud within the challenge window \( T \). This could be modeled based on historical detection times or as a constant if the window is assumed to be adequate for detection.
    \item \textbf{Number of Nodes \( (N) \)}: The total number of nodes that can potentially raise a challenge.
    \item \textbf{Fraudulent Activity \( P(F) \)}: The probability that a transaction or state commitment is fraudulent.
    \item \textbf{Detection Capability \( P(D|F) \)}: The probability of detecting a fraudulent transaction given that it is fraudulent.
    \item \textbf{Participation Rate \( P(R) \)}: The probability that validators are actively monitoring and willing to raise a challenge.
\end{itemize}

\textbf{Probability of a Node Raising a Challenge}: Let \( P(C_i) \) represent the probability that a given node \( i \) raises a challenge. The probability that a single node does not raise a challenge is \( (1 - P(C_i)) \). Therefore, the probability that no node raises a challenge is \( (1 - P(C_i))^N \). Consequently, the probability that at least one node raises a challenge is:

\[
P(\text{at\ least\ one\ challenge}) = 1 - \left(1 - P(C_i)\right)^N
\]

under the assumption that all nodes operates independently.

\vspace{1em}

The probability of a challenge being raised can be modeled as a function of these factors. A simplified form can be expressed as:

\[
P(E) = P(F) \times P(D|F) \times P(T) \times \left[1 - (1 - P(C_i))^N\right]
\]

\vspace{1em}

\textbf{Expected Fraud Rate}: To derive \( P(F) \), consider historical data or theoretical assumptions about the rate of fraudulent transactions. 

\vspace{1em}

\textbf{Detection Probability}: \( P(D|F) \) depends on the effectiveness of the fraud detection mechanisms and the incentives for validators to detect and report fraud.

\vspace{1em}

\textbf{Validator Participation}: \( P(R) \) depends on the number of active validators, their economic incentives, and their responsiveness.

\subsubsection{Example}

\vspace{1em}

\begin{alignat*}{2}
    P(F) &= 0.01 &\quad &\text{(1\% chance a state commitment is fraudulent)} \\
    P(D|F) &= 0.9 &\quad &\text{(90\% chance of detecting fraud if it occurs)} \\
    P(T) &= 1 &\quad &\text{(assuming the challenge window is sufficiently long to detect fraud)} \\
    N &= 100 &\quad &\text{(100 nodes can raise a challenge)} \\
    P(C_i) &= 0.1 &\quad &\text{(10\% chance that a given node raises a challenge)}
\end{alignat*}

\vspace{1em}

Then,

\begin{align*}
    P(E) &= 0.01 \times 0.9 \times 1 \times \left[1 - (1 - 0.1)^{100}\right] \approx 0.0089998
\end{align*}

\vspace{1em}

This result indicates that there is approximately a 0.89\% chance of raising a challenge against a state commitment, which implies that we will achieve the optimal target finality 99.11\% of the time.

\subsubsection{Bonding, Slashing, and Incentives}

Considering the probabilities and their relation to bonding and incentives, we observe the following:

\begin{itemize}
    \item The higher the economic bond of the active operator and the slashing penalties, the lower the probability of fraudulent activity \( P(F) \), as the operator is disincentivized from cheating.
    \item The greater the incentives provided to active challengers (i.e., the security committee), the higher the probabilities of participation \( P(R) \) and fraud detection \( P(D|F) \), ideally approaching \( P(D|F) = 1 \).
    \item Implementing an additional bonding/slashing mechanism for active challengers can help prevent system blockage due to unnecessary challenges.
    \item Additionally, bonding and slashing for active challengers can mitigate the free-rider problem, where challengers confirm state transitions without performing actual checks. Probing with incorrect state diffs can help detect and penalize lazy challengers.
\end{itemize}

With appropriate bonding, incentives, and slashing mechanisms, the system can fine-tune these probabilities to achieve a security level nearly equivalent to that of the base layer. It is worth noting that re-staking protocols such as \cite{jitorestaking2023}, \cite{solayer2023}, and \cite{picasso2024} can help rapidly bootstrap bonding and slashing, thereby enhancing network security.

\section{Future Research}

\subsection{State transitions and Data Availability Sampling }

The trust assumptions discussed in Section \ref{ss:trust-assumptions} rely on active challengers to verify state transitions and raise challenges when necessary. Additionally, the verification of data availability (DA) is crucial, and this can be achieved using methods such as Data Availability Sampling (DAS), as described in \cite{tiny}.

Future work will focus on implementing probabilistic techniques, such as light clients, to verify state transitions and data availability. These approaches aim to reduce the hardware requirements for challenger nodes, thereby enhancing the overall security of the system. The Inter-Blockchain Communication (IBC) protocol \cite{ibc2023} has implemented some promising work in this direction.

\subsection{Validity Fraud Proof}

Finality in fraud-proof systems depends on the length of the challenge period, while in validity-proof systems, it depends on the time needed to generate and verify proofs. Currently, optimistic approaches are more cost-effective and can achieve faster finality using dynamic fraud proofs. However, advancements in validity-proof systems are rapidly reducing the cost and time for proof generation and verification.

As proof systems evolve, there are significant opportunities to improve the efficiency and security of blockchain scaling solutions. While optimistic methods currently offer faster and more cost-effective finality, the ongoing development of validity proofs suggests they may soon match or surpass optimistic approaches. Future research should focus on integrating these advancements with the fast, asynchronous execution of ephemeral rollups.

\section{Conclusion}

The described dynamic fraud-proof mechanism significantly improves the speed of finality under optimal conditions, which can be incentivized through bonding and slashing. It also enables real-time latency when integrated with the Ephemeral Rollups scaling solution for the Solana Virtual Machine. This adaptive mechanism ensures secure and efficient off-chain computation while preserving the integrity and security of the base layer. By addressing key challenges like state fragmentation and censorship attacks, the system provides a scalable, cost-effective solution that can adapt to a range of scenarios.

\clearpage
\printbibliography

\end{document}